\documentclass[a4paper,12pt]{article}

\usepackage{ifpdf}

\newif\ifpdf
\ifx\pdfoutput\undefined
  \pdffalse
\else
  \pdfoutput=1
  \pdftrue
\fi

\RequirePackage{xspace} %
\RequirePackage{subfigure} %
\RequirePackage[centertags]{amsmath} %
\RequirePackage{amssymb}
\RequirePackage{wrapfig} %
\RequirePackage{calc} %
\RequirePackage{ifthen}
\RequirePackage{tabularx} %
\RequirePackage{flafter} %
\RequirePackage{fancyhdr} %

\ifpdf
  \RequirePackage[pdftex]{color}%
  \RequirePackage{colortbl}%
  \RequirePackage{array}%
  \RequirePackage[pdftex]{graphicx}

  \RequirePackage[ pdftex, plainpages = false, pdfpagelabels,
                 pdfpagelayout = useoutlines,
                 bookmarks,
                 breaklinks = true,
                 linktocpage,
                 pagebackref,                      
                 colorlinks = true,
                 linkcolor = blue,
                 urlcolor  = blue,
                 citecolor = blue,
                 anchorcolor = blue,
                 hyperindex = true,
                 hyperfigures
                 ]{hyperref}

\else
  \RequirePackage{color}
  \RequirePackage{colortbl}
   \RequirePackage{array}
  \RequirePackage[dvips]{graphicx}
  \RequirePackage{hyperref}
  \usepackage{rotating}
\fi

\usepackage{makeidx} 
\usepackage{setspace} 
\usepackage{rotating} 
\usepackage{ecltree}
\usepackage{epic}
\usepackage{supertabular}  
\usepackage{color}
\usepackage{exscale}
\usepackage{fontenc}
\usepackage{ifthen}
\usepackage{latexsym}
\usepackage{makeidx}
\usepackage{syntonly}
\usepackage{inputenc}
\usepackage{graphicx}
\usepackage{setspace}
\usepackage{caption2}
\usepackage[english]{babel}
\usepackage[square, comma,numbers,sort&compress]{natbib}
\usepackage{hypernat}
\usepackage{boxedminipage}
\usepackage{framed}
\usepackage{longtable}
\usepackage[all]{hypcap}    
\usepackage{algorithm2e}
\usepackage{algorithmic}
\usepackage{lscape}
\usepackage{pdflscape}

\newcommand{\note}[1]{\marginpar[left]{\singlespace \tiny #1}}
\newcommand{\pois}{Poiseuille}
\newcommand{\Hs}      {\hspace{-0.5cm}} %
\newcommand{\CIF}     {\centering \includegraphics[width=2.7in]} %
\newcommand{\Vmin}    {\vspace{-0.2cm}} %
\newcommand{\codi}    {converging-diverging} %

\renewcommand{\sectionmark}[1]%
      {\markright{\thesection\ #1}} 

\renewcommand{\note}[1]{}

\doublespace 

\title
{ %
\vspace*{3.0cm} \LARGE{\bf Flow of Navier-Stokes Fluids in Converging-Diverging Distensible Tubes} \vspace*{4.0cm} \\
}

\author{Taha Sochi\footnote{University College London, Department of Physics \& Astronomy, Gower Street, London, WC1E 6BT.
Email: t.sochi@ucl.ac.uk.} \vspace*{5.0cm}}

\begin{document}

\maketitle %
\pagenumbering{arabic}

\newpage
\phantomsection \addcontentsline{toc}{section}{Contents} %
\tableofcontents

%

\newpage
\phantomsection \addcontentsline{toc}{section}{Abstract} \noindent
{\noindent \LARGE \bf Abstract} \vspace{0.5cm}\\
\noindent %

We use a method based on the lubrication approximation in conjunction with a residual-based
mass-continuity iterative solution scheme to compute the flow rate and pressure field in
distensible converging-diverging tubes for Navier-Stokes fluids. We employ an analytical formula
derived from a one-dimensional version of the Navier-Stokes equations to describe the underlying
flow model that provides the residual function. This formula correlates the flow rate to the
boundary pressures in straight cylindrical elastic tubes with constant-radius. We validate our
findings by the convergence toward a final solution with fine discretization as well as by
comparison to the \pois-type flow in its convergence toward analytic solutions found earlier in
rigid converging-diverging tubes. We also tested the method on limiting special cases of
cylindrical elastic tubes with constant-radius where the numerical solutions converged to the
expected analytical solutions. The distensible model has also been endorsed by its convergence
toward the rigid \pois-type model with increasing the tube wall stiffness. Lubrication-based
one-dimensional finite element method was also used for verification. In this investigation five
converging-diverging geometries are used for demonstration, validation and as prototypes for
modeling \codi\ geometries in general.


\pagestyle{headings} %
\addtolength{\headheight}{+1.6pt}
\lhead[{Chapter \thechapter \thepage}]%
      {{\bfseries\rightmark}}
\rhead[{\bfseries\leftmark}]%
     {{\bfseries\thepage}} 
\headsep = 1.0cm               

\newpage
\section{Introduction}

The flow of fluids in \codi\ tubes has many scientific, technological and medical applications such
as transportation in porous media, filtration processes, polymer processing, and pathological
stenoses and aneurysms \cite{ShuklaPR1980, Han1981, RuthM1993, DykaarK1996, Sochithesis2007,
ValenciaLRBG2008, SochiFeature2010, GrayM2010, PlappallySFSB2010, BalanBNR2011,
SochiPoreScaleElastic2013, SochiNonNewtBlood2013, SochiBranchFlow2013}. There are many studies
about the flow in \codi\ rigid conduits \cite{ThienGB1985, ThienK1987, BurdetteCAB1989,
JamesTKBP1990, HuzarewiczGC1991, SochiNewtLub2010, SochiPower2011, SochiNavier2013} and distensible
conduits with fixed cross sections \cite{Miekisz1963, RideoutD1967, Heil1998, VajraveluSDP2011,
DescovichPMSW2013, SochiTechnical1D2013, SochiElastic2013} separately as well as many other
different geometries and fluid and conduit mechanical properties \cite{GreidanusDW2011,
GeorgiouK2013, Housiadas2013}. There is also a considerable number of studies on the flow in \codi\
distensible conduits; although large part of which is related to medical applications such as
stenosis modeling \cite{Rao1983, Rao1983b, MisraP1993, TambacaCM2005, BucchiH2013}.

Several methods have been used in the past for investigating and modeling the flow in distensible
\codi\ geometries; the majority are based on the numerical discretization methods such as finite
element and spectral methods although other approaches such as stochastic techniques have also been
employed. However, due to the huge difficulties associating this subject which combines tube wall
deformability with convergence-divergence non-linearities, most of these studies are based on
substantial approximations and modeling compromises. Moreover, they are usually based on very
complex mathematical and computational infrastructures which are not only difficult to implement
and use but also difficult to verify and validate. Also, some of these methods, such as stochastic
techniques, are computationally demanding and hence they may be prohibitive in some cases.
Therefore, simple, reliable and computationally low cost techniques are highly desirable where
analytical solutions are not available due to excessive difficulties or even impossibility of
obtaining such solutions which is the case in most circumstances.

In this paper we propose the use of the lubrication approximation with a residual-based non-linear
solution scheme in association with an analytical expression for the flow of Navier-Stokes fluids
in straight cylindrical elastic tubes with fixed radius to obtain the flow rate and pressure field
in a number of cylindrically-symmetric \codi\ geometries with elastic wall mechanical properties.
The proposed method combines simplicity, robustness and ease of implementation. Moreover, it
produces solutions which are very close to any targeted analytical solutions as the convergence
behavior in the investigated special cases reveals.

Although the proposed method is related to a single distensible tube, it can also be extended to a
network of interconnected distensible tubes with partially or totally \codi\ conduits by
integrating these conduits into the network and giving them a special treatment based on the
proposed method. This approach, can be utilized for example in modeling stenoses and other types of
flow conduits with irregular geometries as part of fluid flow networks in the hemodynamic and
hemorheologic studies and in the filtration investigations.

The method also has a wider validity domain than what may be thought initially with regard to the
deformability characteristics. Despite the fact that in this paper we use a single analytical
expression correlating the flow rate to the boundary pressures for a distensible tube with elastic
mechanical properties, the method can be well adapted to other types of mechanical characteristics,
such as tubes with viscoelastic wall rheology, where different pressure-area constitutive relations
do apply. In fact there is no need even to have an analytical solution for the underlying flow
model that provides the basic flow characterization for the discretized elements of the \codi\
geometries in the lubrication approximation. What is actually needed is only a well defined flow
relation: analytical, or empirical, or even numerical \cite{SochiVariational2013} as long as it is
viable to find the flow in the discretized elements of the lubrication ensemble using such a
relation to correlate the flow rate to the boundary pressures.

There is also no need for the geometry to be of a fixed or regular shape as long as a
characteristic flow can be obtained on the discretized elements, and hence the method can be
applied not only to axi-symmetric geometries with constant-shape and varying cross sectional area
in the flow direction but can also be extended to non-symmetric geometries with irregular and
varying shape along the flow direction if the flow in the deformable discretized elements can be
characterized by a well-defined flow relation. The method can as well be applied to non-straight
flow conduits with and without regular or varying cross sectional shapes such as bending compliant
pipes.

\section{Method} \label{Method}

The flow of Navier-Stokes fluids in a cylindrical tube with a cross sectional area $A$ and length
$L$ assuming a slip-free incompressible laminar axi-symmetric flow with negligible gravitational
body forces and fixed velocity profile is described by the following one-dimensional system of mass
continuity and linear momentum conservation principles

\begin{eqnarray}
\frac{\partial A}{\partial t}+\frac{\partial Q}{\partial x}&=&0\,\,\,\,\,\,\,\,\,\,\,\,\,
t\ge0,\,\,\, x\in[0,L]    \label{NSSystem1} \\
\frac{\partial Q}{\partial t}+\frac{\partial}{\partial x}\left(\frac{\alpha
Q^{2}}{A}\right)+\frac{A}{\rho}\frac{\partial p}{\partial
x}+\kappa\frac{Q}{A}&=&0\,\,\,\,\,\,\,\,\,\,\,\,\, t\ge0,\,\,\, x\in[0,L]     \label{NSSystem2}
\end{eqnarray}

In these two equations, $Q$ is the volumetric flow rate, $t$ is the time, $x$ is the axial
coordinate along the tube length, $\alpha$ is the momentum flux correction factor, $\rho$ is the
fluid mass density, $p$ is the axial pressure which is a function of the axial coordinate, and
$\kappa$ is the viscosity friction coefficient which is usually given by $\kappa
=\frac{2\pi\alpha\nu}{\alpha-1}$ where $\nu$ is the fluid kinematic viscosity defined as the ratio
of the fluid dynamic viscosity $\mu$ to its mass density \cite{BarnardHTV1966, SochiSlip2011,
CostaWM2012, SochiTechnical1D2013, SochiPois1DComp2013, DamianouGM2013, SochiNavier2013}. These
relations are usually supported by a constitutive relation that correlates the pressure to the
cross sectional area in a distensible tube, to close the system in the three variables $A$, $Q$ and
$p$ and hence provide a complete mathematical description for the flow in such conduits.

The correlation between the local pressure and cross sectional area in a compliant tube can be
described by many mathematical constitutive relations depending on the mechanical characterization
of the tube wall and its response to pressure such as being elastic or viscoelastic, and linear or
non-linear. The following is a commonly used pressure-area constitutive elastic relation that
describes such a dependency

\begin{equation}\label{pAEq2}
p=\frac{\beta}{A_{o}}\left(\sqrt{A}-\sqrt{A_{o}}\right)
\end{equation}
where $\beta$ is the tube wall stiffness coefficient which is usually defined by

\begin{equation}\label{beta}
    \beta = \frac{\sqrt{\pi}h_oE}{1-\varsigma^2}
\end{equation}
$A_{o}$ is the reference cross sectional area corresponding to the reference pressure which in this
equation is set to zero for convenience without affecting the generality of the results, $A$ is the
tube cross sectional area at the actual pressure $p$ as opposite to the reference pressure, $h_o$
is the tube wall thickness at the reference pressure, while $E$ and $\varsigma$ are respectively
the Young's elastic modulus and Poisson's ratio of the tube wall. The essence of Equation
\ref{pAEq2} is that the pressure is proportional to the radius growth with a proportionality
stiffness coefficient that is scaled by the reference area. It should be remarked that we assume
here a constant ambient transmural pressure along the axial direction which is set to zero and
hence the reference cross sectional area represents unstressed state where $A_{o}$ is constant
along the axial direction.

Based on the pressure-area relation of Equation \ref{pAEq2}, and using the one-dimensional
Navier-Stokes system of Equations \ref{NSSystem1} and \ref{NSSystem2} for the time-independent flow
by dropping the time terms, the following equation correlating the flow rate $Q$ to the inlet and
outlet boundary areas of an elastic cylindrical tube with a constant unstressed cross sectional
area over its axial direction can be obtained

\begin{equation}\label{QElastic2}
Q=\frac{-\kappa L+\sqrt{\kappa^{2}L^{2}+\frac{4\alpha\beta}{5\rho
A_{o}}\ln\left(A_{in}/A_{ou}\right)\left(A_{in}^{5/2}-A_{ou}^{5/2}\right)}}{2\alpha\ln\left(A_{in}/A_{ou}\right)}
\end{equation}
where $A_{in}$ and $A_{ou}$ are the tube cross sectional area at the inlet and outlet respectively
such that $A_{in}>A_{ou}$. This relation, which in essence correlates the flow rate to the boundary
pressures, has been previously \cite{SochiElastic2013} derived and validated by a finite element
scheme.

The residual-based lubrication approach, which is proposed in the present paper to find the
pressure field and flow rate in \codi\ distensible tubes, starts by discretizing the tube in the
axial direction into ring-like elements. Each one of these elements is approximated as a single
tube with a constant radius, which averages the inlet and outlet radii of the element, to which
Equation \ref{QElastic2} applies. A system of non-linear equations based on the mass continuity
residual and boundary conditions is then formed.

For a tube discretized into ($N-1$) elements, there are $N$ nodes: two boundaries and ($N-2$)
interior nodes. Each one of these nodes has a well-defined axial pressure value according to the
one-dimensional formulation. Also for the interior nodes, and due to the incompressibility of the
flow, the total sum of the volumetric flow rate, signed ($+/-$) according to its direction with
respect to the node, is zero due to the absence of sources and sinks, and hence ($N-2$) residual
functions which describe the net flow at the interior nodes can be formed. This is associated with
two given boundary conditions for the inlet and outlet boundary nodes to form $N$ equations.

A standard method for solving such a system is to use an iterative non-linear simultaneous solution
scheme such as Newton-Raphson method where an initial guess for the interior nodal pressures is
proposed and used in conjunction with the Jacobian matrix of the system to find the pressure
perturbation vector which is then used to adjust the pressure values and repeat this process until
a convergence criterion based on the size of the residual norm is reached. The process is based on
iterative solving of the following equation

\begin{equation}\label{JDpR1}
\mathbf{J}\Delta\mathbf{p}=-\mathbf{r}
\end{equation}
where $\mathbf{J}$ is the Jacobian matrix, $\mathbf{p}$ is the vector of variables which represent
the pressure values at the boundary and interior nodes, and $\mathbf{r}$ is the vector of residuals
which, for the interior nodes, is based on the continuity of the volumetric flow rate as given by

\begin{equation}
f_{j}=\sum_{i=1}^{m}Q_{i}=0
\end{equation}
where $m$ is the number of discretized elements connected to node $j$ which is two in this case,
and $Q_{i}$ is the signed volumetric flow rate in element $i$ as characterized by Equation
\ref{QElastic2}. Equation \ref{JDpR1} is then solved in each iteration for $\Delta\mathbf{p}$ which
is then used to update $\mathbf{p}$. The convergence will be declared when the norm of the residual
vector, $\mathbf{r}$, becomes within a predefined error tolerance. More details about this solution
scheme can be found in \cite{SochiTechnical1D2013, SochiPoreScaleElastic2013}.

\section{Implementation and Results}

\begin{figure}[!h]
\centering{}
\includegraphics[scale=0.95]{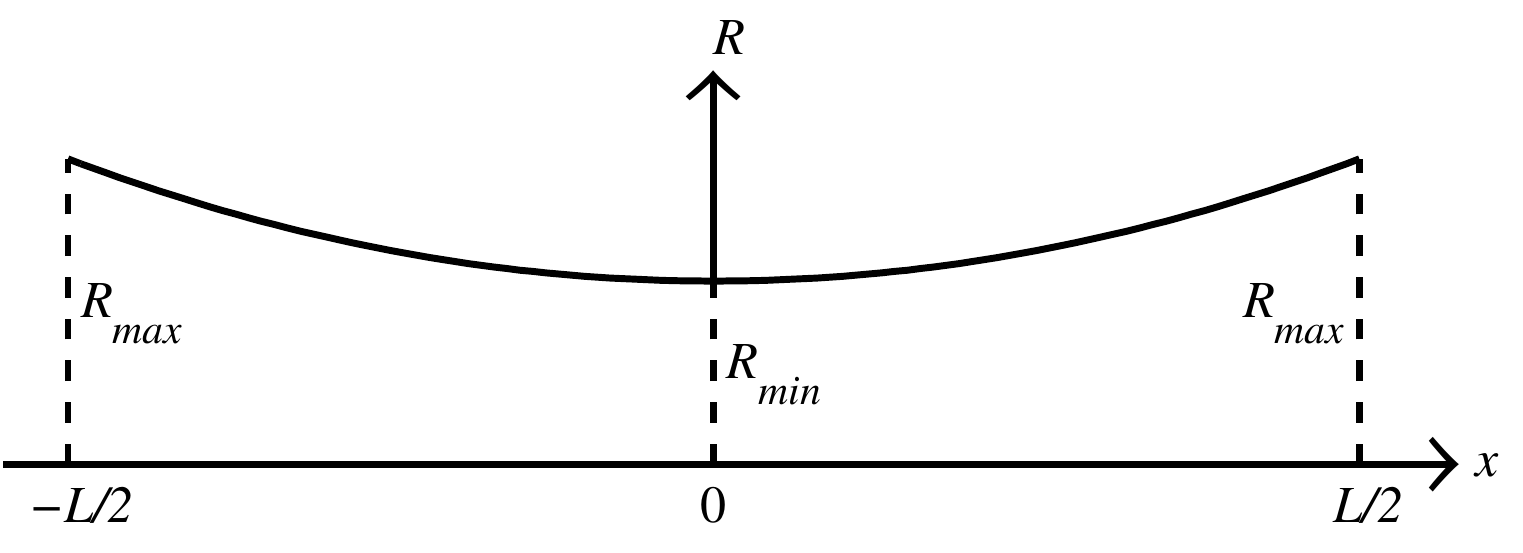}
\caption{Generic \codi\ tube profile demonstrating the coordinate system setting for the
correlation between the axial coordinate $x$ and the tube radius $R$ used in Table
\ref{GeometryTable}.} \label{GenericProfile}
\end{figure}

\begin{figure}[!h]
\centering{}
\includegraphics[scale=0.75]{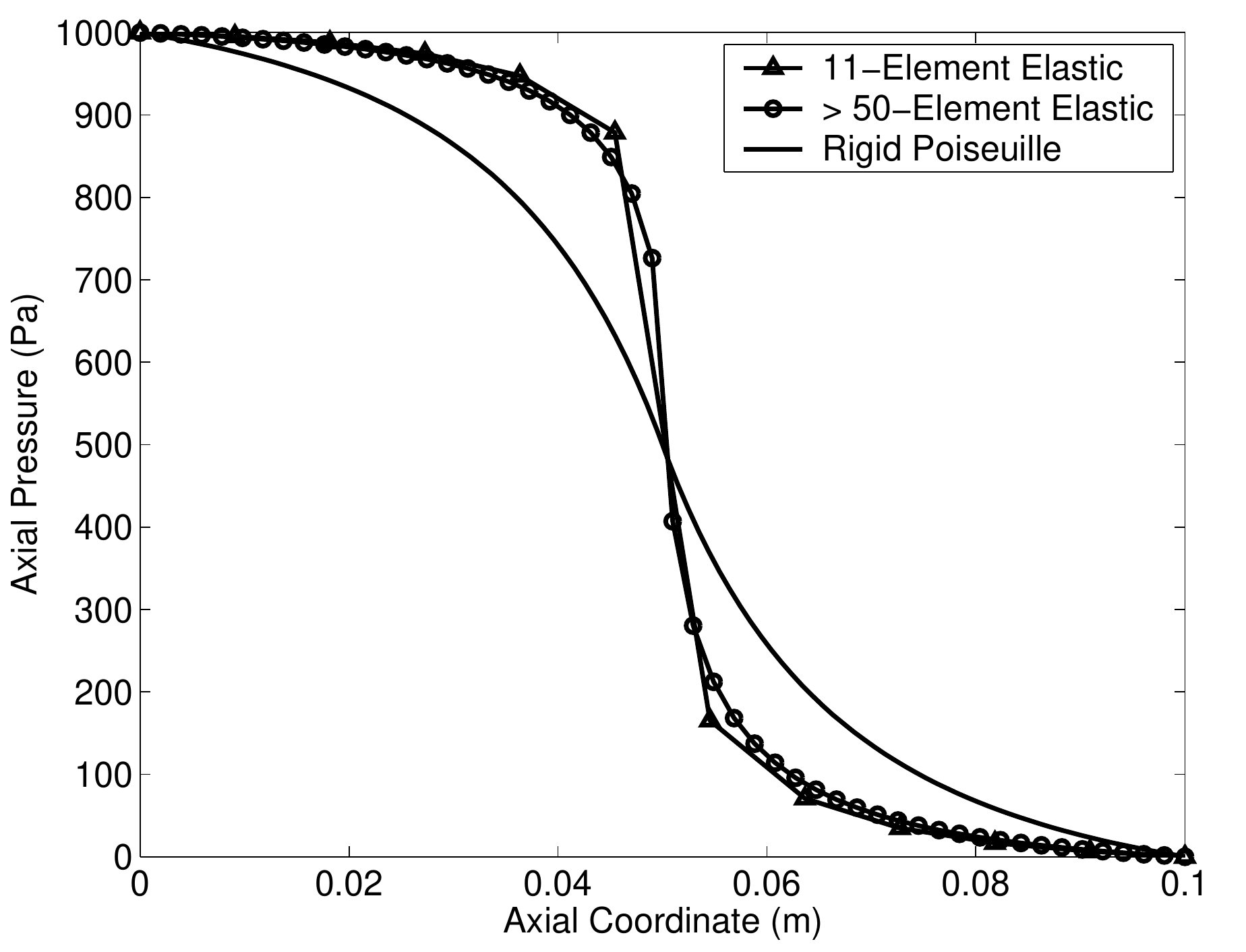}
\caption{Axial pressure as a function of axial coordinate for a \codi\ elastic tube with conic
geometry having $L=0.1$~m, $R_{min}=0.005$~m, $R_{max}=0.01$~m, and $\beta=236.3$~Pa.m. The fluid
properties are: $\rho=1000$~kg.m$^{-3}$ and $\mu=0.01$~Pa.s while the inlet and outlet pressures
are: $p_{i}=1000$~Pa and $p_{o}=0.0$~Pa. The \pois-type flow uses a rigid tube with the same
unstressed geometry and the same $\mu$ and boundary pressures. The converged flow rate for the
elastic Navier-Stokes and rigid \pois-type flows are respectively:
$Q_{e}=0.000255889$~m$^3$.s$^{-1}$ and $Q_{r}=0.000842805$~m$^3$.s$^{-1}$ while the analytic flow
rate for the rigid tube as obtained from the first equation in Table \ref{QTable} is
$Q_{a}=0.000841498$~m$^3$.s$^{-1}$.} \label{ConicQ}
\end{figure}

\begin{figure}[!h]
\centering{}
\includegraphics[scale=0.75]{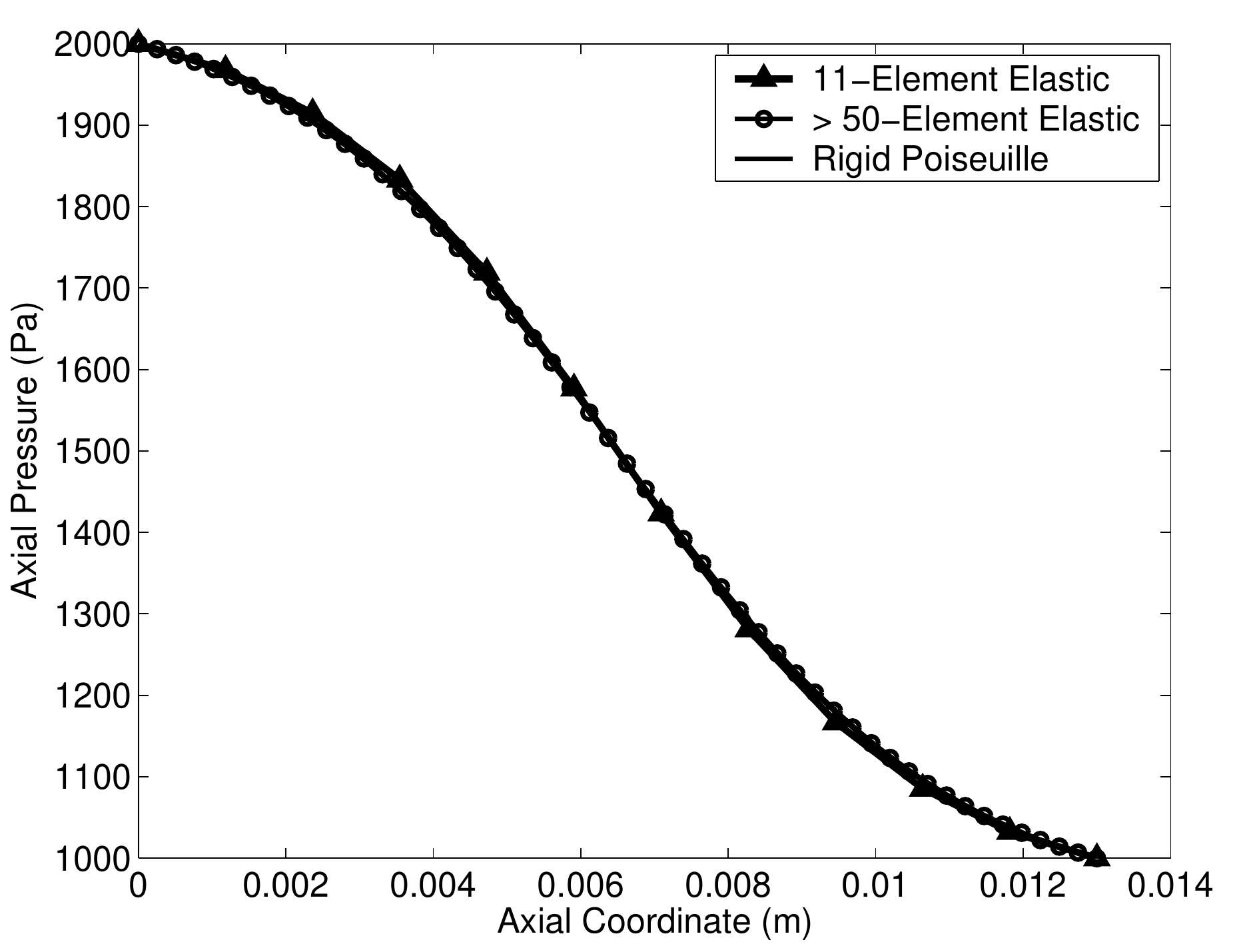}
\caption{Axial pressure as a function of axial coordinate for a \codi\ elastic tube with parabolic
geometry having $L=0.013$~m, $R_{min}=0.0017$~m, $R_{max}=0.0025$~m, and $\beta=28059.0$~Pa.m. The
fluid properties are: $\rho=1100$~kg.m$^{-3}$ and $\mu=0.006$~Pa.s while the inlet and outlet
pressures are: $p_{i}=2000$~Pa and $p_{o}=1000$~Pa. The \pois-type flow uses a rigid tube with the
same unstressed geometry and the same $\mu$ and boundary pressures. The converged flow rate for the
elastic Navier-Stokes and rigid \pois-type flows are respectively: $Q_{e}=6.58209\times
10^{-5}$~m$^3$.s$^{-1}$ and $Q_{r}=6.62929\times 10^{-5}$~m$^3$.s$^{-1}$ while the analytic flow
rate for the rigid tube as obtained from the second equation in Table \ref{QTable} is
$Q_{a}=6.62051\times 10^{-5}$~m$^3$.s$^{-1}$.} \label{ParabolicQ}
\end{figure}

\begin{figure}[!h]
\centering{}
\includegraphics[scale=0.75]{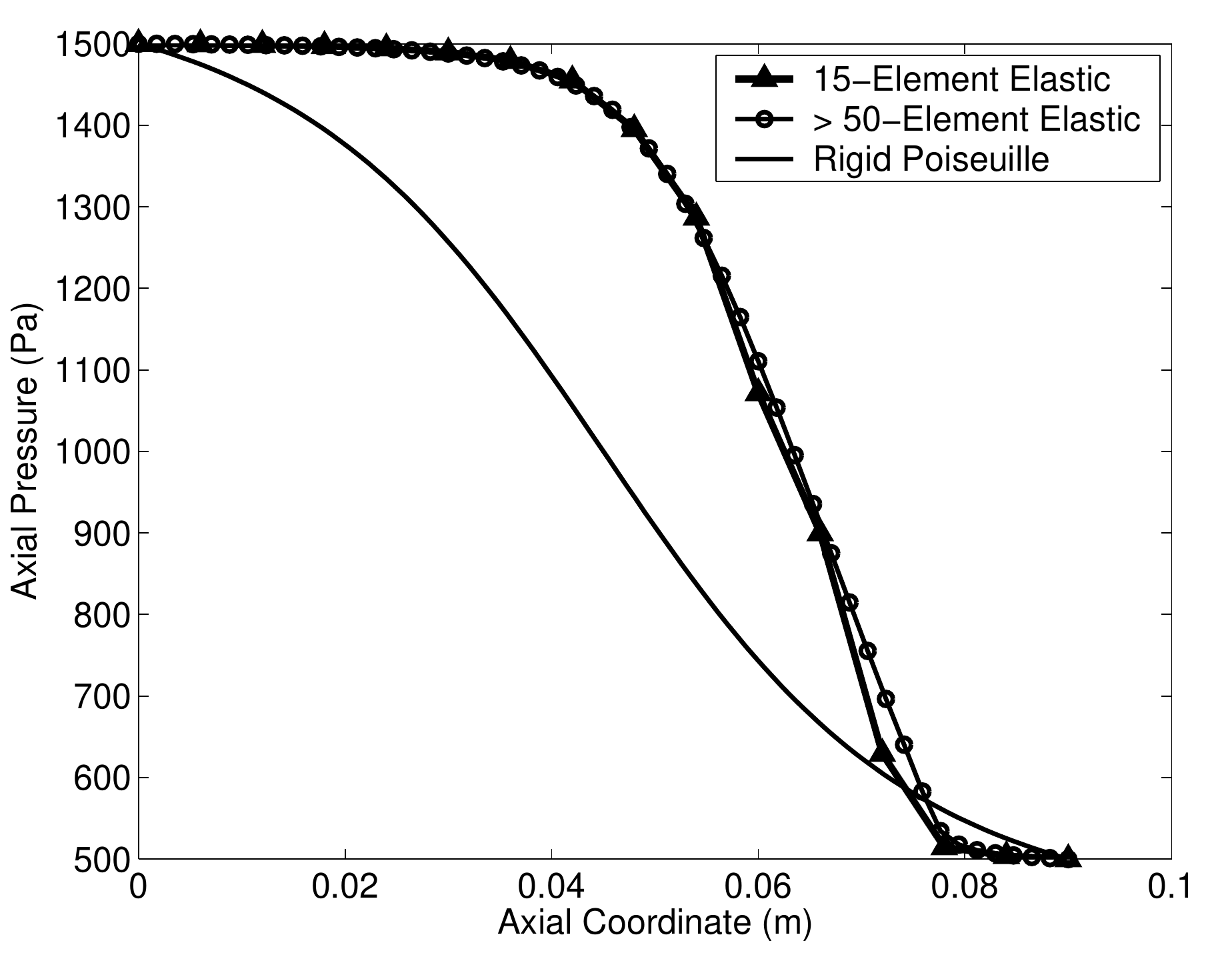}
\caption{Axial pressure as a function of axial coordinate for a \codi\ elastic tube with hyperbolic
geometry having $L=0.09$~m, $R_{min}=0.004$~m, $R_{max}=0.006$~m, and $\beta=23.6$~Pa.m. The fluid
properties are: $\rho=800$~kg.m$^{-3}$ and $\mu=0.002$~Pa.s while the inlet and outlet pressures
are: $p_{i}=1500$~Pa and $p_{o}=500$~Pa. The \pois-type flow uses a rigid tube with the same
unstressed geometry and the same $\mu$ and boundary pressures. The converged flow rate for the
elastic Navier-Stokes and rigid \pois-type flows are respectively:
$Q_{e}=0.000147335$~m$^3$.s$^{-1}$ and $Q_{r}=0.000934645$~m$^3$.s$^{-1}$ while the analytic flow
rate for the rigid tube as obtained from the third equation in Table \ref{QTable} is
$Q_{a}=0.000933394$~m$^3$.s$^{-1}$.} \label{HyperbolicQ}
\end{figure}

\begin{figure}[!h]
\centering{}
\includegraphics[scale=0.75]{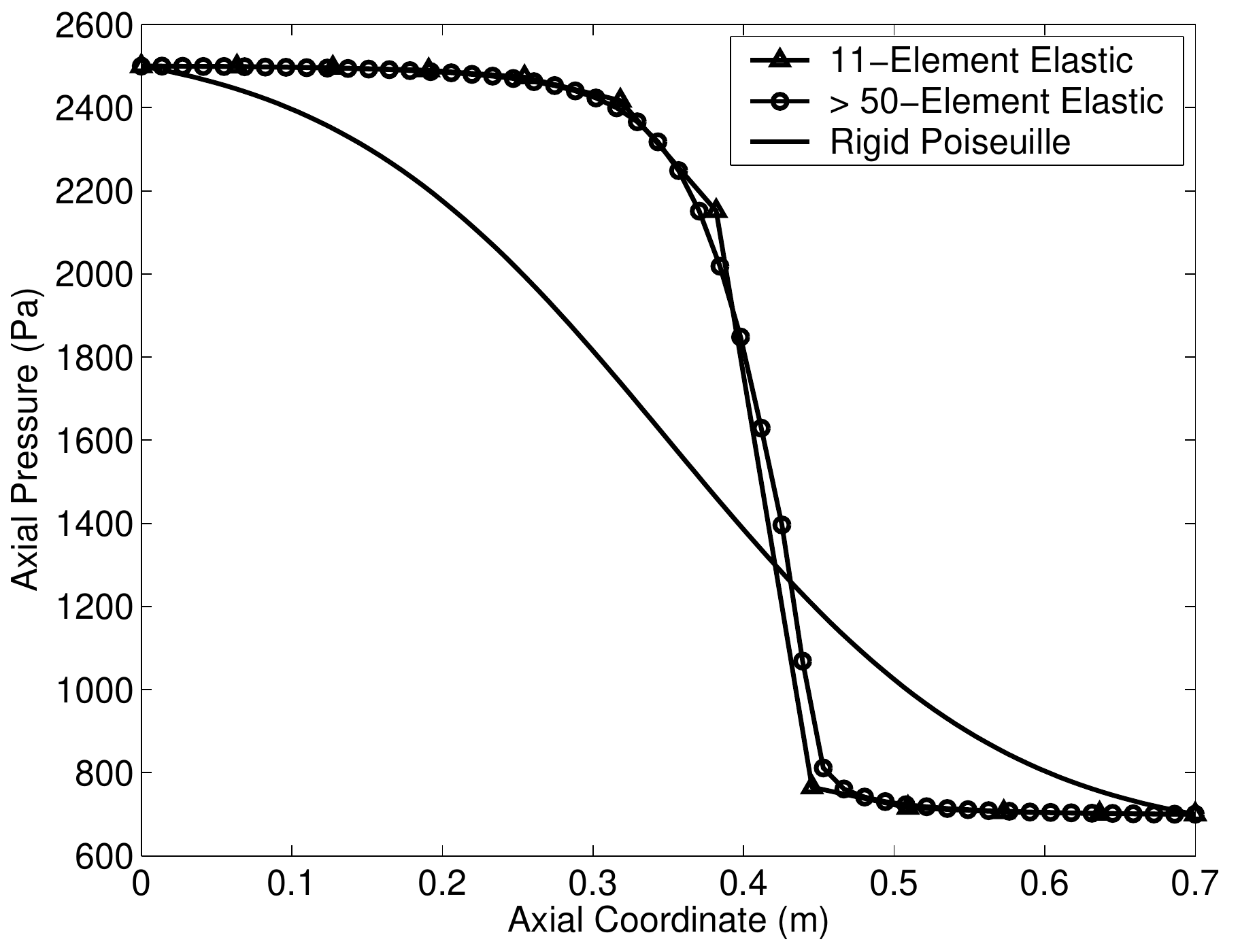}
\caption{Axial pressure as a function of axial coordinate for a \codi\ elastic tube with hyperbolic
cosine geometry having $L=0.7$~m, $R_{min}=0.05$~m, $R_{max}=0.08$~m, and $\beta=3889.4$~Pa.m. The
fluid properties are: $\rho=700$~kg.m$^{-3}$ and $\mu=0.0075$~Pa.s while the inlet and outlet
pressures are: $p_{i}=2500$~Pa and $p_{o}=700$~Pa. The \pois-type flow uses a rigid tube with the
same unstressed geometry and the same $\mu$ and boundary pressures. The converged flow rate for the
elastic Navier-Stokes and rigid \pois-type flows are respectively: $Q_{e}=0.0427687$~m$^3$.s$^{-1}$
and $Q_{r}=1.4184$~m$^3$.s$^{-1}$ while the analytic flow rate for the rigid tube as obtained from
the fourth equation in Table \ref{QTable} is $Q_{a}=1.416296$~m$^3$.s$^{-1}$.} \label{CoshQ}
\end{figure}

\begin{figure}[!h]
\centering{}
\includegraphics[scale=0.75]{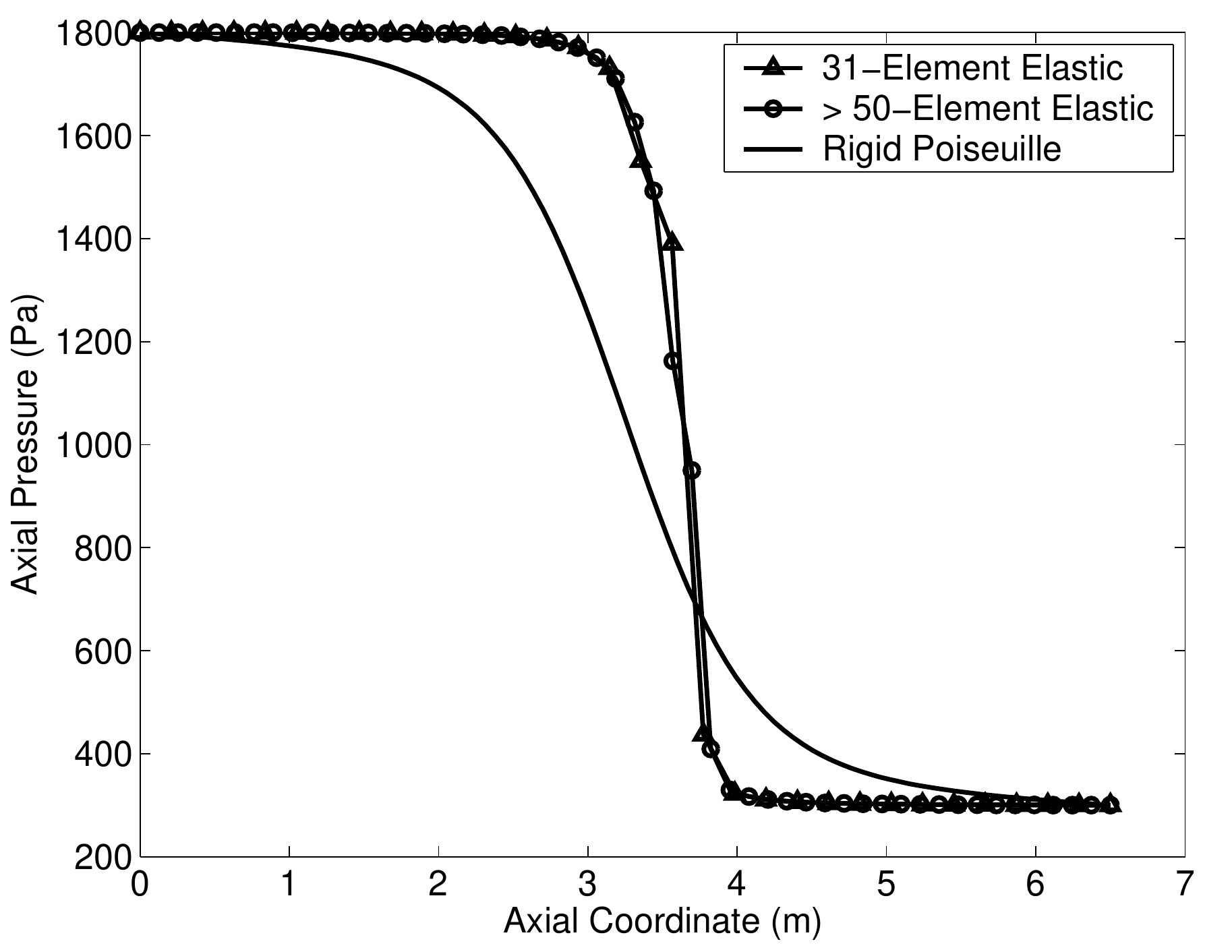}
\caption{Axial pressure as a function of axial coordinate for a \codi\ elastic tube with sinusoidal
geometry having $L=6.5$~m, $R_{min}=0.2$~m, $R_{max}=0.5$~m, $\beta=5064.2$~Pa.m. The fluid
properties are: $\rho=900$~kg.m$^{-3}$ and $\mu=0.06$~Pa.s while the inlet and outlet pressures
are: $p_{i}=1800$~Pa and $p_{o}=300$~Pa. The \pois-type flow uses a rigid tube with the same
unstressed geometry and the same $\mu$ and boundary pressures. The converged flow rate for the
elastic Navier-Stokes and rigid \pois-type flows are respectively: $Q_{e}=0.396769$~m$^3$.s$^{-1}$
and $Q_{r}=8.74955$~m$^3$.s$^{-1}$ while the analytic flow rate for the rigid tube as obtained from
the fifth equation in Table \ref{QTable} is $Q_{a}=8.73370$~m$^3$.s$^{-1}$.} \label{SinusoidQ}
\end{figure}

The proposed residual-based lubrication method was implemented in a computer code and flow
solutions were obtained for an extensive range of fluid, flow and tube characterizations such as
fluid viscosity, flow profile, and tube mechanical properties. Five regular \codi\ axi-symmetric
tube geometries were used in the current investigation; representative graphic images of these
geometries are shown in Figure \ref{Geometries}, while the mathematical relations that describe the
dependency of the tube radius, $R$, on the tube axial coordinate, $x$, for these geometries are
given in Table \ref{GeometryTable}. A generic \codi\ tube profile demonstrating the setting of the
coordinate system for the $R$--$x$ correlation, as used in Table \ref{GeometryTable}, is shown in
Figure \ref{GenericProfile}. These geometries have been used previously \cite{SochiPower2011,
SochiNavier2013} to find flow relations for Newtonian and power law fluids in rigid tubes. A
representative sample of the flow solutions on distensible \codi\ tubes are also given in Figures
\ref{ConicQ}-\ref{SinusoidQ}.


\begin{figure} [!h]
\centering %
\subfigure[Conic]%
{\begin{minipage}[b]{0.5\textwidth} \CIF {g/Conic}
\end{minipage}}
\Hs %
\subfigure[Parabolic]%
{\begin{minipage}[b]{0.5\textwidth} \CIF {g/Parabolic}
\end{minipage}} \Vmin

%
\centering %
\subfigure[Hyperbolic]%
{\begin{minipage}[b]{0.5\textwidth} \CIF {g/Hyperbolic}
\end{minipage}}
\Hs %
\subfigure[Hyperbolic Cosine]%
{\begin{minipage}[b]{0.5\textwidth} \CIF {g/HyperbolicCosine}
\end{minipage}} \Vmin

%
\centering %
\subfigure[Sinusoidal]%
{\begin{minipage}[b]{0.5\textwidth} \CIF {g/Sinusoidal}
\end{minipage}}
\caption{Converging-Diverging tube geometries used in the current investigation.
\label{Geometries}}
\end{figure}


\begin{table} [!h]
\caption{The equations describing the  dependency of the tube radius $R$ on the tube axial
coordinate $x$ for the five converging-diverging geometries used in the current investigation. In
all these relations $-\frac{L}{2}\le x\le\frac{L}{2}$ and $R_{min}<R_{max}$ where $R_{min}$ is the
tube minimum radius at $x=0$ and $R_{max}$ is the tube maximum radius at $x=\pm \frac{L}{2}$ as
demonstrated in Figure \ref{GenericProfile}.} \label{GeometryTable} \vspace{0.2cm} \centering
\begin{tabular}{ll}
 \hline
Geometry & \hspace{2cm}$R(x)$ \\
 \hline
 Conic & $R_{min}+\frac{2(R_{max}-R_{min})}{L}|x|$ \vspace{0.3cm} \\
 Parabolic & $R_{min}+\left(\frac{2}{L}\right)^{2}(R_{max}-R_{min})x^{2}$ \vspace{0.3cm} \\
 Hyperbolic & $\sqrt{R_{min}^{2}+\left(\frac{2}{L}\right)^{2}(R_{max}^{2}-R_{min}^{2})x^{2}}$ \vspace{0.3cm} \\
 Hyperbolic Cosine \hspace{3.5cm} & $R_{min}\cosh\left[\frac{2}{L}\mathrm{arccosh}\left(\frac{R_{max}}{R_{min}}\right)x\right]$ \vspace{0.3cm} \\
 Sinusoidal & $\left(\frac{R_{max}+R_{min}}{2}\right)-\left(\frac{R_{max}-R_{min}}{2}\right)\cos\left(\frac{2\pi x}{L}\right)$ \vspace{0.3cm} \\
\hline
\end{tabular}
\end{table}

In all flow simulations, including the ones shown in Figures \ref{ConicQ}-\ref{SinusoidQ}, we used
a range of evenly-divided discretization meshes to observe the convergence behavior of the solution
with respect to mesh refinement. In all cases we noticed an obvious trend of convergence with
improved meshing toward a final solution that does not tangibly improve with further mesh
refinement. We also used in these flow simulations a rigid conduit flow model with the same
geometry and fluid and flow properties where the flow in the rigid discretized elements was modeled
by \pois\ equation. The purpose of this use of the rigid model is to assess the solution scheme and
test its convergence to the correct solution because for \pois-type flow with rigid geometries we
have analytical solutions, given in Table \ref{QTable}, that correlate the flow rate to the
pressure drop. \pois-type solutions can also provide a qualitative indicator of the sensibility of
the distensible solutions; for instance we expect the deviation between the two solutions to
decrease with increasing the stiffness of the elastic tube. In all cases the correct quantitative
values and qualitative trends have been verified.

\begin{table} [!h]
\caption{The equations describing the dependency of the flow rate $Q$ on the pressure drop $\Delta
p$ for the rigid tubes with the five \codi\ geometries of Table \ref{GeometryTable}. These
relations were previously \cite{SochiNavier2013} derived and validated.} \label{QTable} \vspace{0.2cm} \centering {
\begin{tabular}{ll}
\hline
Geometry & \hspace{4cm}$Q(\Delta p)$ \vspace{0.2cm} \\
\hline
Conic &  $\frac{3\pi^{2}\Delta p}{\kappa\rho L}\left(\frac{R_{min}^{3}R_{max}^{3}}{R_{min}^{2}+R_{min}R_{max}+R_{max}^{2}}\right)$ \vspace{0.3cm} \\
Parabolic & $\frac{2\pi^{2}\Delta p}{\kappa\rho L}\left(\frac{1}{3R_{min}R_{max}^{3}}+\frac{5}{12R_{min}^{2}R_{max}^{2}}+\frac{5}{8R_{min}^{3}R_{max}}+\frac{5\arctan\left(\sqrt{\frac{R_{max}-R_{min}}{R_{min}}}\right)}{8R_{min}^{7/2}\sqrt{R_{max}-R_{min}}}\right)^{-1}$ \vspace{0.3cm} \\
Hyperbolic & $\frac{2\pi^{2}\Delta p}{\kappa\rho L}\left(\frac{1}{R_{min}^{2}R_{max}^{2}}+\frac{\arctan\left(\sqrt{\frac{R_{max}^{2}-R_{min}^{2}}{R_{min}^{2}}}\right)}{R_{min}^{3}\sqrt{R_{max}^{2}-R_{min}^{2}}}\right)^{-1}$ \vspace{0.3cm} \\
Cosh & $\frac{3\pi^{2}\Delta p}{\kappa\rho L}\left(\frac{\mathrm{arccosh}\left(\frac{R_{max}}{R_{min}}\right)R_{min}^{4}}{\tanh\left(\mathrm{arccosh}\left(\frac{R_{max}}{R_{min}}\right)\right)\left[\mathrm{sech}^{2}\left(\mathrm{arccosh}\left(\frac{R_{max}}{R_{min}}\right)\right)+2\right]}\right)$ \vspace{0.3cm} \\
Sinusoidal & $\frac{16\pi^{2}\Delta p}{\kappa\rho L}\left(\frac{(R_{max}R_{min})^{7/2}}{2(R_{max}+R_{min})^{3}+3(R_{max}+R_{min})(R_{max}-R_{min})^{2}}\right)$ \vspace{0.3cm} \\
\hline
\end{tabular}}
\end{table}

Each one of Figures \ref{ConicQ}-\ref{SinusoidQ} shows a sample of the numeric solutions for two
sample meshes used for the distensible flow geometry alongside the converged \pois-type solution
for the given fluid and tube parameters. The reason for showing two meshes for the distensible
geometry is to demonstrate the convergence behavior with mesh refinement. In all cases, virtually
identical solutions were obtained with meshes finer than the finest one shown in these figures.

It should be remarked that in all the distensible flow simulations shown in Figures
\ref{ConicQ}-\ref{SinusoidQ} we used $\alpha=4/3$ to match the rigid \pois-type flow profile
\cite{SochiNavier2013} which we used, as indicated already, as a test case. However, for the
purpose of testing and validating the distensible model in general we also used an extensive range
of values greater than and less than $4/3$ for $\alpha$ without observing incorrect convergence or
convergence difficulties. In fact using values other than $\alpha=4/3$ makes the convergence easier
in many cases \cite{SochiPoreScaleElastic2013}.

\begin{figure} [!h]
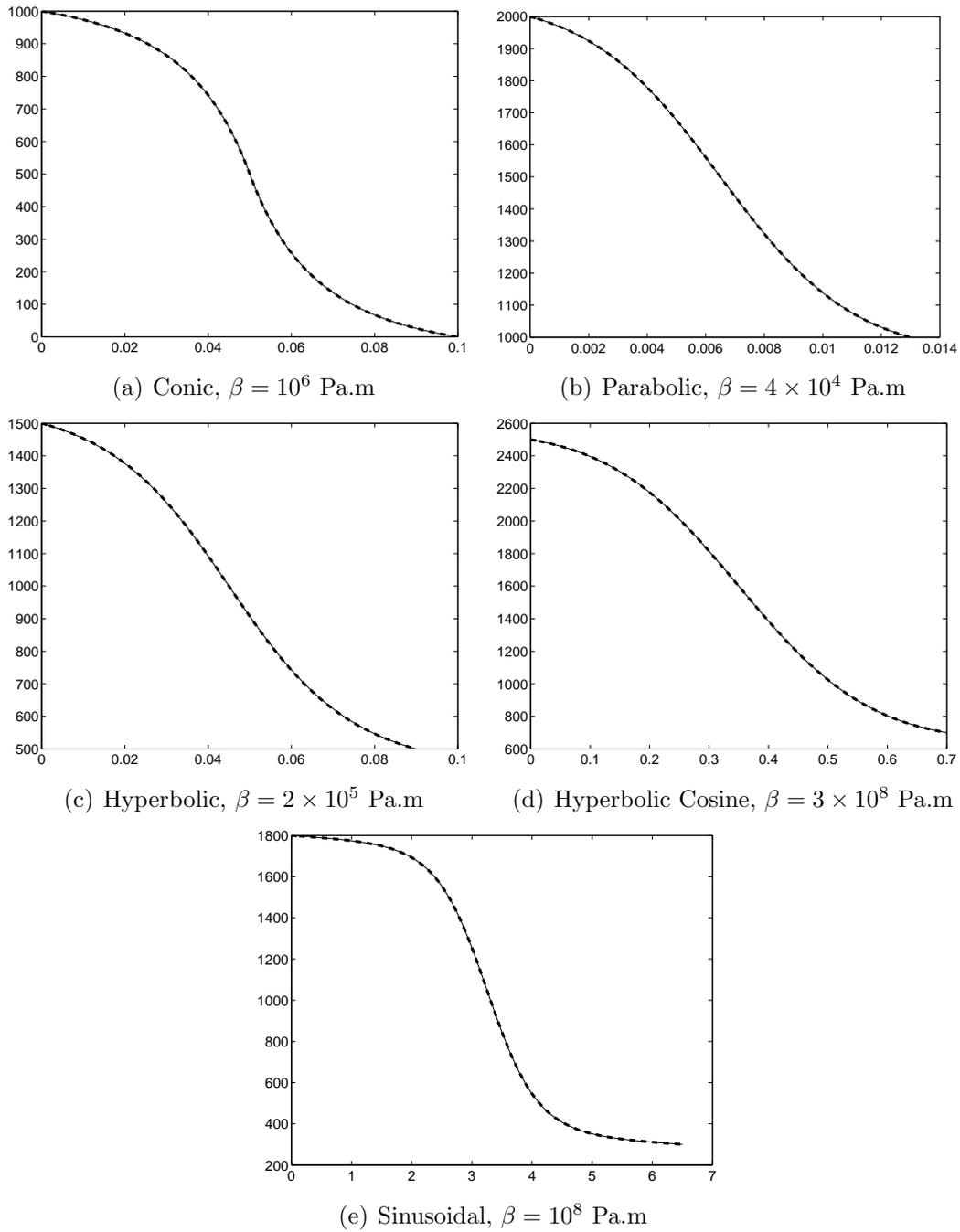

\centering %
\subfigure[Conic, $\beta=10^6$ Pa.m]%
{\begin{minipage}[b]{0.5\textwidth} \CIF {g/StiffConic}
\end{minipage}}
\Hs %
\subfigure[Parabolic, $\beta=4\times10^4$ Pa.m]%
{\begin{minipage}[b]{0.5\textwidth} \CIF {g/StiffParabolic}
\end{minipage}} \Vmin

%
\centering %
\subfigure[Hyperbolic, $\beta=2\times10^5$ Pa.m]%
{\begin{minipage}[b]{0.5\textwidth} \CIF {g/StiffHyperbolic}
\end{minipage}}
\Hs %
\subfigure[Hyperbolic Cosine, $\beta=3\times10^8$ Pa.m]%
{\begin{minipage}[b]{0.5\textwidth} \CIF {g/StiffHyperbolicCosine}
\end{minipage}} \Vmin

%
\centering %
\subfigure[Sinusoidal, $\beta=10^8$ Pa.m]%
{\begin{minipage}[b]{0.5\textwidth} \CIF {g/StiffSinusoidal}
\end{minipage}}
\caption{Comparing the converged \pois-type rigid tube flow (solid) to the converged elastic tube
flow with high wall stiffness of the given $\beta$ (dashed) for the five examples of Figures
\ref{ConicQ}-\ref{SinusoidQ}. In all the five sub-figures, the vertical axis represents the axial
pressure in pascals while the horizontal axis represents the tube axial coordinate in meters. The
converged numeric flow rate in each case for the rigid and elastic models is virtually identical to
the corresponding \pois-type analytic flow rate given in Figures \ref{ConicQ}-\ref{SinusoidQ}.
\label{StiffFig}}
\end{figure}

An interesting feature that can be seen in Figure \ref{ParabolicQ} is that all the pressure profile
curves are almost identical as well as the flow rates. The reason is that, due to the high tube
stiffness used in this example, the distensible tube solution converged to the rigid tube
\pois-type solution. A more detailed comparison between the \pois-type rigid tube flow and the
Navier-Stokes one-dimensional elastic tube flow with high stiffness is shown in Figure
\ref{StiffFig} where the results of Figures \ref{ConicQ}-\ref{SinusoidQ} are reproduced using the
same fluid, flow and tube parameters but with high tube stiffness by using large $\beta$'s. As seen
in Figure \ref{StiffFig} the elastic tube flow converges almost identically to the \pois-type rigid
tube flow with increasing the tube wall stiffness in all cases. This sensible and correct trend can
be regarded as another verification and validation for the residual-based method and the related
computer code. Similar results have also been obtained in \cite{SochiPois1DComp2013} in comparing
the rigid and distensible models for the flow in networks of interconnected straight cylindrical
tubes. More detailed comparisons between the rigid and distensible one-dimensional flow models can
be found in the aforementioned reference.

It should be remarked that the critical value of $\beta$ at which the distensible flow solution
converges to the rigid flow solution depends on several factors such as the fluid and flow
parameters as well as the geometry of the tube and the pressure field regime characterized by the
applied boundary conditions at the inlet and outlet where their size and the magnitude of their
difference play a decisive role. Another remark is that the shape of the pressure profile curve is
highly dependent on the geometric factors such as $\frac{L}{R_{min}}$, $\frac{L}{R_{max}}$, and
$\frac{R_{min}}{R_{max}}$ ratios. It also depends on the fluid and tube mechanical properties, such
as fluid viscosity and tube wall stiffness, and the magnitude of pressure at the inlet and outlet
boundaries.

The opposite to what in Figure \ref{ParabolicQ} can be seen in Figure \ref{HyperbolicQ} for the
hyperbolic geometry where we used very low stiffness and hence the elastic model deviated largely
from the rigid model. This also affected the dependency of convergence rate on discretization where
the discrepancy between the solutions of the coarse and fine meshes was more substantial than in
the other cases for similar coarse and fine meshes. In general, the deviation between the rigid and
distensible flow models is maximized by reducing the stiffness, and hence increasing the tube
distensibility, while other parameters are kept fixed.

Another interesting feature is that in the flow solution of Figure \ref{CoshQ} there is a big
difference between the flow rate of the elastic and rigid tubes. This can be explained largely by
the significant deviation from linearity due to the large values of the inlet and outlet boundary
pressures, as well as the large size of their difference, with a relatively low stiffness. This
indicates that the rigid tube flow model is not a suitable approximation for simulating and
analyzing the flow in distensible tubes and networks, as it has been done for instance in some
hemodynamic studies. More detailed discussions about this issue can be found in
\cite{SochiPois1DComp2013}.

\begin{figure} [!h]
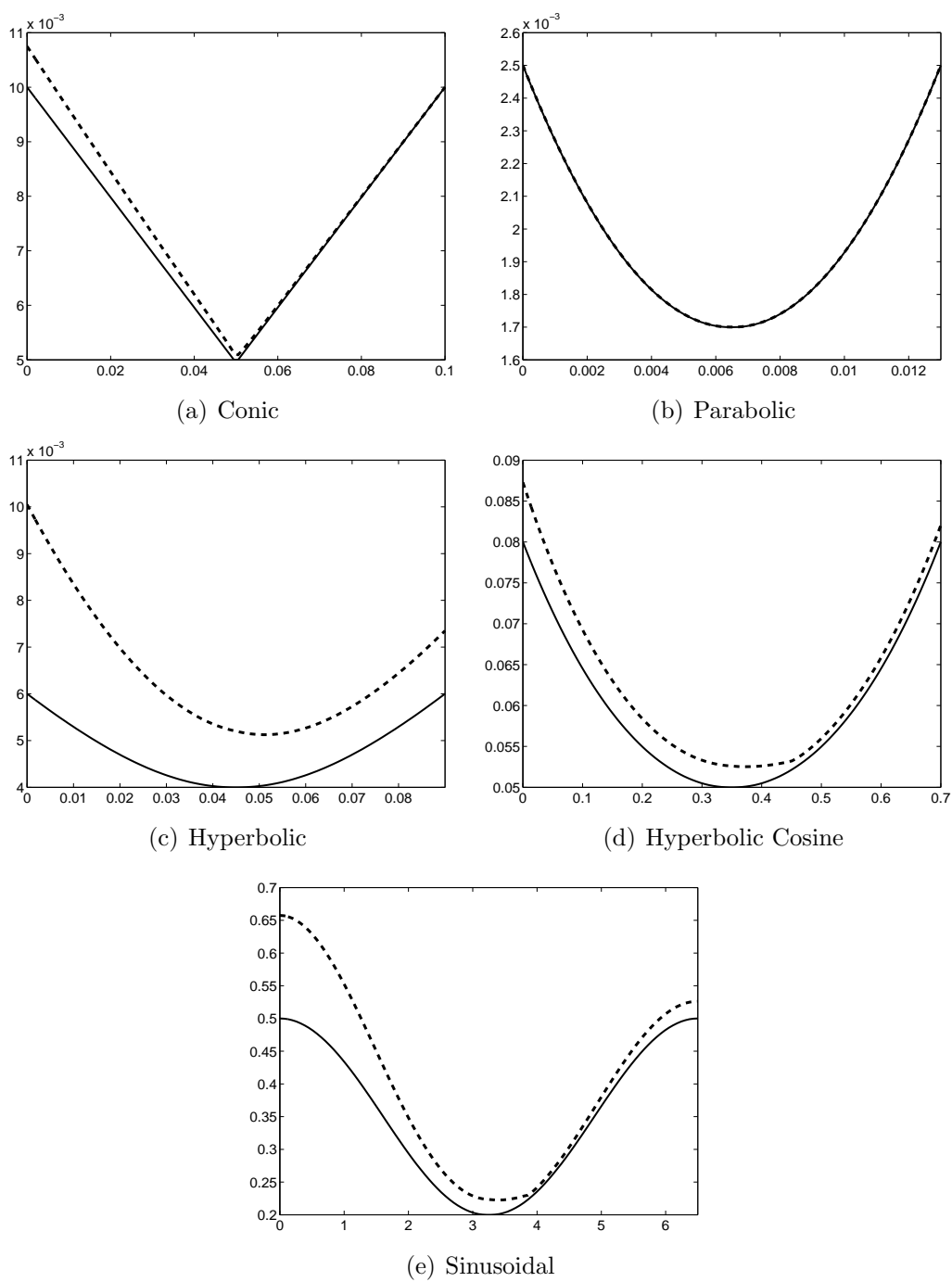

\centering %
\subfigure[Conic]%
{\begin{minipage}[b]{0.5\textwidth} \CIF {g/ProfileConic}
\end{minipage}}
\Hs %
\subfigure[Parabolic]%
{\begin{minipage}[b]{0.5\textwidth} \CIF {g/ProfileParabolic}
\end{minipage}} \Vmin

%
\centering %
\subfigure[Hyperbolic]%
{\begin{minipage}[b]{0.5\textwidth} \CIF {g/ProfileHyperbolic}
\end{minipage}}
\Hs %
\subfigure[Hyperbolic Cosine]%
{\begin{minipage}[b]{0.5\textwidth} \CIF {g/ProfileHyperbolicCosine}
\end{minipage}} \Vmin

%
\centering %
\subfigure[Sinusoidal]%
{\begin{minipage}[b]{0.5\textwidth} \CIF {g/ProfileSinusoidal}
\end{minipage}}
\caption{Comparing the elastic tube unstressed radius (solid) to the stressed radius (dashed) as a
function of the tube axial coordinate for the five examples of Figures
\ref{ConicQ}-\ref{SinusoidQ}. In all the five sub-figures, the vertical axis represents the tube
radius in meters and the horizontal axis represents the tube axial coordinate in meters as well.
\label{ProfileFig}}
\end{figure}

In Figure \ref{ProfileFig} we draw the geometric profile of the elastic tube for the stressed and
unstressed states for the five examples of Figures \ref{ConicQ}-\ref{SinusoidQ} where we plot the
tube radius versus its axial coordinate for these two states. As seen, these plots show another
sensible qualitative trend in these results and hence provide further endorsement to the
residual-based method. It is needless to say that in Figures \ref{ConicQ}-\ref{ProfileFig} the
inlet boundary is at $x=0$ while the outlet boundary is at the other end.

\section{Tests and Validations}

We used several metrics to validate the residual-based method and check our computer code and flow
solutions. First, we did extensive tests on distensible cylindrical tubes with fixed radius using
different fluid, flow and tube parameters where the method produced results identical to the
analytical flow solutions given by Equation \ref{QElastic2}. Although this test is based on a
simple limiting case and hence it may be regarded as trivial, it provides sufficient validation for
the basic approach and the reliability of the code. We also investigated the convergence behavior,
outlined in the previous section, as a function of discretization; in all cases it was observed
that the residual-based method converges to a final solution with the use of finer meshes where it
eventually stabilizes without tangible change in the solution with more mesh refinement. This
convergence behavior is a strong qualitative indicator for the accuracy of the method and the
reliability of the code. As indicated previously, we used evenly-divided regular meshes in all
simulations.

We also used the discretized \pois-type flow in the same \codi\ geometry but with rigid wall
mechanical characteristics to validate the solutions, as discussed in the previous section. As
seen, we observed in all cases the convergence of the \pois-type solutions on using reasonably fine
meshes to the analytical solutions with errors that are comparable to the machine precision and
hence are negligible as they are intrinsic to any machine-based numerical method. Since the elastic
and rigid models are based on the same mathematical and computational infrastructure, the
convergence of the rigid flow model to the correct analytical solution can be regarded as an
indirect endorsement to the elastic model. The convergence of the elastic model solution to the
verified rigid model solution with increasing tube wall stiffness is another indirect support for
the elastic model as it demonstrates its sensible behavior.

As another way of test and validation, we produced a sample of lubrication-based one-dimensional
finite element solutions which are obtained by discretizing the \codi\ distensible geometries and
applying the pressure continuity, rather than the Bernoulli energy conservation principle, as a
coupling condition at the nodal interfaces \cite{SochiTechnical1D2013, SochiBranchFlow2013} to
match the assumptions of the residual-based method which couples the discretized elements by the
continuity of pressure condition \cite{SochiPoreScaleElastic2013}. The finite element results were
very similar to the residual-based results although the convergence behavior was generally
different. The residual-based method has a better convergence behavior in general although this is
highly dependent on coding technical issues and implementation specificities and hence cannot be
generalized.

With regard to the comparison between the residual-based and finite element methods, they have very
similar theoretical infrastructure as they are both based on the same formulation of the
one-dimensional Navier-Stokes flow. In fact the residual-based method is a modified version of the
previously proposed \cite{SochiPoreScaleElastic2013} pore-scale network modeling method for the
flow of Navier-Stokes fluids in networks of interconnected distensible tubes by extending the
concept of a network to serially-connected tubes with varying radii which represent the discretized
elements of the \codi\ tubes. Hence the agreement between the residual-based and finite element
methods may not be regarded as an entirely independent validation method and that is why we did not
do detailed validation by the lubrication-based one-dimensional finite element.

\section{Comparisons}

As indicated previously, the advantages of the residual-based method in comparison to other methods
include simplicity, ease of implementation, low computational costs, and reliability of solutions
which are comparable in their accuracy to any intended analytical solutions based on the given
assumptions, as the investigated limiting cases like rigid and fixed-radius tubes have revealed.
These advantages also apply for the residual-based method in comparison to the lubrication-based
one-dimensional finite element method plus a better overall convergence behavior. The biggest
advantage of the finite element method, however, is its applicability to the transient
time-dependent flow and more suitability for probing other flow-related one-dimensional transport
phenomena such as the reflection and propagation of pressure waves. Therefore, the
lubrication-based one-dimensional finite element could be the method of choice for investigating
transient flow and wave propagation in distensible geometries until proper modifications are
introduced on the residual-based method to extend it to these modalities. More details about the
comparison between the residual-based and finite element methods can be found in
\cite{SochiPoreScaleElastic2013}.

The residual-based method, as indicated earlier, can also be used for irregular flow conduits in
general with cross sections that vary in size and shape and even without \codi\ feature and
regardless of being cylindrically axi-symmetric as long as an analytical, or empirical, or even
numerical \cite{SochiVariational2013} relation between the boundary pressures and flow rate on a
straight geometry with a similar cross sectional shape does exist. Therefore it can be safely
claimed that the residual-based method has a wider applicability range than many other methods
whose explicit or implicit underlying assumptions apply only to restricted types of conduit
geometry.

With regard to convergence, each numerical method has its own characteristic convergence behavior
which depends on many factors such as the utilized numerical solvers and their underlying
mathematical and computational theory, the nature of the physical problem, the employed convergence
support techniques, coding technicalities, and so on. Hence it is not easy to make a definite
comparison for the convergence behavior between different numerical methods. However, we can say
that the residual-based method has in general a better rate and speed of convergence in comparison
to other commonly-used numerical methods. More details about convergence issues and convergence
enhancement techniques can be found in \cite{SochiPoreScaleElastic2013}.

On the other hand, the residual-based method has a number of limitations based on its underlying
physical assumptions, as stated in section \ref{Method}, as well as limitations rooted in its
one-dimensional nature that restricts its applicability to modeling axially-dependent flow
phenomena and hence excludes phenomena related to other types of dependency. However, most of these
limitations are shared by other comparable methods.

\section{Conclusions}

A simple and reliable method based on the lubrication approximation in conjunction with a
non-linear simultaneous solution scheme based on the continuity of pressure and volumetric flow
rate with an analytical solution correlating the flow rate to the boundary pressures in straight
cylindrical elastic tubes with constant radius is used in this paper to find the flow rate and
pressure field in distensible tubes with \codi\ shapes. Five \codi\ axi-symmetric geometries were
used for demonstrating the applicability of the method and assessing its merit.

The method is validated by its convergence behavior with finer discretization as well as comparing
the equivalent \pois-based flow to the analytical solutions which were obtained and validated
previously. A sample of lubrication-based one-dimensional finite element solutions have also been
obtained and compared to the residual-based solutions; these results show very good agreement. The
method was also tested on limiting cases of elastic cylindrical tubes with fixed radius, where it
produced results identical to the analytical solutions, as well as the convergence to the
established rigid tube flow with increasing tube wall stiffness.

The method can be extended to geometries other than cylindrically axi-symmetric \codi\ shapes as
long as a flow characterization relation can be provided for the discretized elements; whether
analytical or empirical or even numerical. The method can also be extended beyond the use in
computing the flow in single tubes to computing the flow in networks of interconnected distensible
conduits which are, totally or partially, characterized by having \codi\ geometries, or variable
cross sectional shapes or curving structure in the flow direction to be more general.

Many industrial and medical applications, such as material processing and stenosis modeling, can
benefit from this approach which is easy to implement and integrate with other flow modeling
techniques. Moreover, it produces highly accurate solutions with low computational costs. An
initial investigation indicates that its convergence behavior is generally superior to that of the
traditional numerical techniques such as the one-dimensional finite element especially with the use
of convergence enhancement techniques.

\clearpage
\phantomsection \addcontentsline{toc}{section}{Nomenclature} %
{\noindent \LARGE \bf Nomenclature} \vspace{0.5cm}

\begin{supertabular}{ll}
$\alpha$                &   correction factor for axial momentum flux \\
$\beta$                 &   stiffness coefficient in the pressure-area relation \\
$\kappa$                &   viscosity friction coefficient \\
$\mu$                   &   fluid dynamic viscosity \\
$\nu$                   &   fluid kinematic viscosity \\
$\rho$                  &   fluid mass density \\
$\varsigma$             &   Poisson's ratio of tube wall \\
\\
$A$                     &   tube cross sectional area at actual pressure \\
$A_{in}$                &   tube cross sectional area at inlet \\
$A_o$                   &   tube cross sectional area at reference pressure \\
$A_{ou}$                &   tube cross sectional area at outlet \\
$E$                     &   Young's elastic modulus of the tube wall \\
$f$                     &   flow continuity residual function \\
$h_o$                   &   tube wall thickness at reference pressure \\
$\mathbf{J}$            &   Jacobian matrix \\
$L$                     &   tube length \\
$N$                     &   number of discretized tube nodes \\
$p$                     &   pressure \\
$\mathbf{p}$            &  pressure vector \\
$p_i$                     &  inlet pressure \\
$p_o$                     &  outlet pressure \\
$\Delta p$              & pressure drop \\
$\Delta\mathbf{p}$      &   pressure perturbation vector \\
$Q$                     &   volumetric flow rate \\
$Q_{a}$                 &   analytic flow rate for rigid tube \\
$Q_{e}$                 &   numeric flow rate for elastic tube \\
$Q_{r}$                 &   numeric flow rate for rigid tube \\
$\mathbf{r}$             &  residual vector \\
$R$                     &   tube radius \\
$R_{max}$               &   maximum unstressed tube radius \\
$R_{min}$               &   minimum unstressed tube radius \\
$t$                     &   time \\
$x$                     &   tube axial coordinate \\

\end{supertabular}

\clearpage
\phantomsection \addcontentsline{toc}{section}{References} %
\bibliographystyle{unsrt}

\end{document}